\documentclass[aps,prb,showpacs,twocolumn,amsmath,amssymb]{revtex4-2}

\usepackage{amsmath}    
\usepackage{amssymb}
\usepackage{graphicx}   
\usepackage{graphics}
\usepackage{dcolumn}
\usepackage{bm}

\usepackage{multirow}

\usepackage{latexsym}
\usepackage{amssymb}
\usepackage{amsmath}

\usepackage{color}
\usepackage{epstopdf}

\usepackage[normalem]{ulem}

\begin{document}
\title{The influence of antiferromagnetic spin cantings on the magnetic helix pitch in cubic helimagnets }
\author{Viacheslav A. Chizhikov\footnote{email: chizhikov@crys.ras.ru} and Vladimir E. Dmitrienko\footnote{email: dmitrien@crys.ras.ru}}
\affiliation{A.V. Shubnikov Institute of Crystallography, FSRC ``Crystallography
and Photonics'' RAS, Leninskiy Prospekt 59, 119333, Moscow, Russia}

\pacs{}

\begin{abstract} 
In cubic helimagnets MnSi and Cu$_2$OSeO$_3$ with their nearly isotropic magnetic properties, the magnetic structure undergoes helical deformation, which is almost completely determined by the helicoid wavenumber $k = \mathcal{D} / \mathcal{J}$, where magnetization field stiffness $\mathcal{J}$ is associated with isotropic spin exchange, and $\mathcal{D}$ is a pseudoscalar value characterizing the antisymmetric Dzyaloshinskii--Moriya (DM) interaction. While the wavenumber can be measured directly in a diffraction experiment, the values of $\mathcal{J}$ and $\mathcal{D}$ can be calculated from the constants of pair spin interactions, which enter as parameters into the Heisenberg energy. However, the available analytical expression for $\mathcal{D}$, which is of the first order in the spin-orbit coupling (SOC), has significant problems with accuracy. Here we show that hardly observable distortions of the magnetic structure, namely the antiferromagnetic spin cantings, can significantly change the constant $\mathcal{D}$ in the next approximation in SOC, thus affecting the wavenumber of magnetic helicoids. The obtained analytical expressions agree with the results of numerical simulation of magnetic helices in Cu$_2$OSeO$_3$ to within a few percent.
\end{abstract}
\maketitle


\section{Introduction}
\label{sec:intro}

The helical order in Nature is found in a variety of systems, from condensed matter physics to biology \cite{Kornyshev2007}, and serves as an important example of primitive self-organization. This self-organization can create rather complex structures, such as blue phases in cholesteric liquid crystals \cite{Belyakov1985,Wright1989} or skyrmion textures in chiral cubic magnets MnSi, MnGe, Fe$_{1-x}$Co$_{x} $Si, Cu$_2$OSeO$_3$, etc. \cite{Bogdanov1989,BogdanovPSSB1994,BogdanovJMMM1994,Rossler2006,Stishov2011,Borisov2020}. At the same time, there is a remarkable similarity in the physics of such different systems \cite{Tewari2006,Binz2006,Hamann2011}. Nevertheless, in each individual case, its own microscopic mechanisms are responsible for the emergence of the helical order, and significant efforts are required to find them. In particular, it is necessary to understand how local interactions in a system affect its final structure. For example, the first question is: what determines the period and direction of spirals? On the other hand, the description of the microscopic particularities is important both for applications such as spintronics and multiferroics \cite{Pyatakov2012,Mostovoy2006,PyatakovUFN2015} and for fundamental problems such as the topological Hall effect \cite{Kanazawa2011}.

Since the discovery of the chiral magnetic properties of MnSi in 1976 \cite{Ishikawa1976,Motoya1976}, the phenomenological theory based on the Ginzburg--Landau (GL) free energy has been used to describe and predict twisted magnetic structures \cite{Bak1980,Nakanishi1980}. However, this approach, which uses our knowledge of the symmetry of a physical system, cannot tell how large the values of the free energy coefficients are and how they relate to the spin coupling parameters. Microscopic theories, such as the Heisenberg model of a classical ferromagnet with a spin-orbit term proposed by Dzyaloshinskii and Moriya \cite{Dzyaloshinskii1957,Dzyaloshinskii1958,Moriya1960a,Moriya1960b}. in turn, have the disadvantage that they are difficult to use in any analytical calculations. Nevertheless, despite some doubts about the correctness of the Heisenberg model for itinerant magnets, it is often used for numerical simulations \cite{Hamann2011,Hopkinson2009,Chen2015,Mueller2020,Hall2021}. For this model, it is necessary to know the parameters of spin interactions: constants $J_{ij}$ of the isotropic exchange between $i$th and $j$th atoms, and pseudovectors $\mathbf{D}_{ij}$ of the antisymmetric Dzyaloshinskii--Moriya (DM) exchange. These parameters can be obtained in two ways: from a comparison of theoretical predictions with experimental data and from {\it ab initio} calculations \cite{GongXiang2012,Janson2014}. Using the  Ruderman--Kittel--Kasuya--Yoshida theory that is more relevant for itinerant magnets allows one to calculate only the magnetization stiffness $\mathcal{J}$, but not the DM interaction parameter $\mathcal{D}$ \cite{Kashin2018,Chizhikov2013}.

These two approaches, microscopic and phenomenological, have been used for many years to describe the same magnetic structures, complementing each other well. However, for a correct transition from one model to another, even for such a simple case as MnSi, it took a long time. The transition consists in expressing the constants entering the GL energy of the continuum model in terms of the parameters $J_{ij}$, $\mathbf{D}_{ij}$ of the Heisenberg model. The reverse transition is obviously ambiguous due to the large number of microscopic parameters. In Refs.~\cite{Dmitrienko2012,Chizhikov2012}, such a transition was carried out for the first time for MnSi crystals in the nearest neighbor approximation. An important side result of this work was the prediction of antiferromagnetic spin cantings, which are an inherent feature of the magnetic structure of helimagnets. Further, the approach was first extended by taking into account interactions with next-nearest neighbors in MnSi \cite{Chizhikov2013} and, finally, generalized to the case of other cubic helimagnets, including Cu$_2$OSeO$_3$ \cite{Chizhikov2015}. For the multiferroic Cu$_2$OSeO$_3$, the spin cantings play an additional important role, connecting its magnetic and ferroelectric properties \cite{Chizhikov2017}. An alternative mathematical approach describing the transition between the microscopic and phenomenological models for crystals with the $B20$ structure (FeGe) is developed in Refs.~\cite{Grytsiuk2019,Grytsiuk2021}.

Previously, when passing from the microscopic model to the continuum one for cubic helimagnets, only contributions to the energy up to the second order in spin-orbit coupling (SOC), inclusive, were taken into account. However, the accuracy of the transition remains unexplored, and there is reason to believe that this approximation is insufficient for a correct calculation of the wavenumber of spin helicoids. In this paper, we will calculate the third-order SOC contributions to the energy and show that they provide a good approximation for calculating wavenumbers. In Sec.~\ref{sec:status}, we describe the current state of the problem of transition from a microscopic model to a continuum one and show that the Keffer rule \cite{Keffer1969} for the DM interaction requires an increase in the calculation accuracy. In Sec.~\ref{sec:transition}, the GL energy for cubic helimagnets is derived in the third approximation in SOC. In Sec.~\ref{sec:example}, the obtained expressions are used to calculate the wavenumber of helicoids for the ferrimagnet Cu$_2$OSeO$_3$, and the comparison with the results of numerical simulation turns out to be quite satisfactory. Section~\ref{sec:discussion} discusses the effect of a strong magnetic field on the wavenumber, and the relationship between the DM interaction and magnetic anisotropy.

\section{Discrete--continuum transition: \\ A preview}
\label{sec:status}
Before proceeding to a consistent description of the transition from the microscopic model of a cubic helimagnet to its phenomenological model, we briefly describe the current state of affairs in this area. The problem of correspondence between discrete (atomistic) and continuum descriptions of matter has always been in the branches of physics that study fluids and solids, whether it is the theory of elasticity, fluid mechanics or electrodynamics of continuous media. The transition from a discrete model to a continuum one is often ambiguous, and the reverse transition has no physical meaning. In particular, the transition from discrete spins ordered in the magnetic lattice of a crystal to a continuous magnetization field is not unambiguous and depends on the method of spatial averaging of the magnetic moments, in the case when the magnetization changes along the crystal.

Neglecting the crystal field associated with the local anisotropy of the positions of magnetic atoms, the energy of a Heisenberg magnet with antisymmetric DM exchange can be written in the following form:
\begin{equation}
\label{eq:EHeisenberg}
\begin{split}
E = \frac12 \sum_{i,j} \left( - J_{ij} \mathbf{s}_i \cdot \mathbf{s}_j + \mathbf{D}_{ij} [\mathbf{s}_i \times \mathbf{s}_j] \right) \\ - \mathbf{H} \cdot \sum_i g_i \mu_\mathrm{B} \mathbf{s}_i ,
\end{split}
\end{equation}
where $\mathbf{s}_i$ is the classical spin ($|\mathbf{s}_i| = 1$) related to the magnetic moment of the $i$th atom as $\mathbf{m}_i = g_i \mu_\mathrm{B} \mathbf{s}_i$, $g_i$ is the spin $g$-factor, $\mathbf{H}$ is an external magnetic field, $J_{ij}$ and $\mathbf{D}_{ij}$ are the parameters of the isotropic spin exchange and the DM interaction, respectively. The first sum is taken over all pairs of interacting magnetic atoms, and each pair is included in the sum twice ($i,j \equiv j,i$), which requires the coefficient $\frac12$.

The permutation relations $J_{ji} = J_{ij}$, $\mathbf{D}_{ji} = - \mathbf{D}_{ij}$ are obvious from Eq.~(\ref{eq:EHeisenberg}). In addition, for equivalent bonds, the isotropic exchange constants are the same, and the DM vectors are related by the symmetry operations of the crystal point group. Note that for crystals without an inversion center, which we are interested in, all these symmetry elements are rotations, while in the case of point groups with mirror symmetry, $\mathbf{D}_{ij}$ should behave like pseudovectors.

The DM interaction has relativistic spin-orbit nature, so the small parameter $(v/c)^2$ arises naturally, where $v$ is a characteristic velocity of electrons in the crystal. Note that the quadratic part of the energy (\ref{eq:EHeisenberg}) should also include the anisotropic term $\mathbf{s}_i \cdot \hat{\mathcal{P}}_{ij} \cdot \mathbf{s}_j$, where $\hat{\mathcal{P}}_{ij}$ is a symmetric traceless tensor. This tensor is also related to the DM interaction, e.g. from Refs.~\cite{Hopkinson2009,Shekhtman1992,Yildirim1995} $\hat{\mathcal{P}}_{ij} = - (\mathbf{D}_{ij} \otimes \mathbf{D}_ {ij} - \frac13 \mathbf{D}_{ij}^2 \hat{I}) / 2 / J_{ij}$. However, since $\hat{\mathcal{P}}_{ij}$ is of the second order in SOC and contributes to the energy starting from the fourth-order terms, we will neglect it here. Note that the spin-orbit interaction also corrects the isotropic exchange, $\Delta J_{ij} = \mathbf{D}_{ij}^2 / 12 / J_{ij}$. Without loss of generality, we can assume that this correction is already included in $J_{ij}$.

The Heisenberg energy (\ref{eq:EHeisenberg}) allows one to simulate magnetic structures in terms of lattice models. However, it is often convenient to use the continuum approximation based on the expression for the energy density as a function of the magnetization field, which varies smoothly in the crystal. A naive transition from one model to another is carried out by introducing a slowly varying unimodular field $\boldsymbol{\mu}(\mathbf{r})$ ($|\boldsymbol{\mu}| = 1$), which coincides with classical spins in all atomic positions: $\mathbf{s}_i = \boldsymbol{\mu}(\mathbf{r}_i)$, $\mathbf{s}_j = \boldsymbol{\mu}(\mathbf{r}_j)$. For small interatomic distances
\begin{equation}
\label{eq:muTaylor}
\boldsymbol{\mu}(\mathbf{r}_j) \approx \boldsymbol{\mu}(\mathbf{r}_i) + (\mathbf{r}_{ij} \cdot \boldsymbol{\nabla}) \boldsymbol{\mu}(\mathbf{r}_i) + \frac12 (\mathbf{r}_{ij} \cdot \boldsymbol{\nabla})^2 \boldsymbol{\mu}(\mathbf{r}_i) ,
\end{equation}
where $\mathbf{r}_{ij} \equiv \mathbf{r}_j - \mathbf{r}_i$, and thus we can pass from Eq.~(\ref{eq:EHeisenberg}) to a GL-like energy, which is a functional of the field $\boldsymbol{\mu}(\mathbf{r})$.

For cubic helimagnets of the MnSi and Cu$_2$OSeO$_3$ type, the energy density containing terms of the second order in SOC has the form
\begin{equation}
\label{eq:ELandau}
\mathcal{E} = \frac12 \mathcal{J} \frac{\partial \boldsymbol{\mu}}{\partial r_\alpha} \cdot \frac{\partial \boldsymbol{\mu}}{\partial r_\alpha} + \mathcal{D} \boldsymbol{\mu} \cdot \mathrm{curl} \boldsymbol{\mu} - \mathbf{H} \cdot M_0 \boldsymbol{\mu} ,
\end{equation}
where $M_0 = \sum_{i}^\prime g_i \mu_\mathrm{B}$ is the saturation magnetization, and the magnetic structure stiffness and  DM constant are calculated as follows \cite{Chizhikov2015}:
\begin{equation}
\label{eq:J&D}
\mathcal{J} = \frac16 \left.\sum_{i,j}\right.^\prime J_{ij} r_{ij}^2 , \phantom{x} \mathcal{D} = - \frac16 \left.\sum_{i,j}\right.^\prime \mathbf{D}_{ij} \cdot \mathbf{r}_{ij} ,
\end{equation}
where the prime means that the summation is carried out over a unit cell. Note that, due to the high symmetry of cubic crystals, Eq.~(\ref{eq:ELandau}) does not contain terms related to crystal anisotropy. The first anisotropic terms in the energy of cubic helimagnet is the fourth order in SOC.

Eqs.~(\ref{eq:J&D}) reveal the link between the parameters of two models, discrete and continuum, describing the same magnetic crystal. Besides, it is obvious that these expressions cannot be correct. Indeed, the formulas for $\mathcal{J}$ and $\mathcal{D}$ contain interatomic distances $\mathbf{r}_{ij}$, which are not present in the initial model based on the Heisenberg energy (\ref{eq:EHeisenberg}). A complete analysis of this problem was done in Ref.~\cite{Chizhikov2013}. It has been shown that, at the microscopic level, the spin structure looks more complicated than the naive transition from Eq.~(\ref{eq:EHeisenberg}) to Eq.~(\ref{eq:ELandau}) suggests. Two types of microscopic particularities have been predicted to break the smoothness of the field $\boldsymbol{\mu}(\mathbf{r})$. First, instead of one magnetization field, one should consider several fields corresponding to the magnetic sublattices of the crystal. It is the neglect of phase shifts between the magnetic sublattices that leads to incorrect expressions (\ref{eq:J&D}). Secondly, there are antiferromagnetic cantings of neighboring spins, which do not disappear even in a structure polarized by a strong magnetic field \cite{Dmitrienko2012}. In Ref.~\cite{Chizhikov2013}, a formal technique has been suggested that takes into account the contribution of phase shifts. For this, we propose to use some ideal positions (``exchange coordinates'') $\tilde{\mathbf{r}}_i$ of magnetic atoms in the unit cell, for which the phase shifts vanish. The exchange coordinates are defined as functions of the isotropic exchange constants $J_{ij}$ and have the same symmetry (Wyckoff position) as the real coordinates of the atoms. The technique allows saving Eqs.~(\ref{eq:J&D}) for the parameters of the continuum model, where the exchange distances $\tilde{\mathbf{r}}_{ij} \equiv \tilde{\mathbf{r}}_j - \tilde{\mathbf{r}}_i$ are substituted instead of the real interatomic distances $\mathbf{r}_{ij}$. As for the antiferromagnetic cantings, they give a constant contribution to the energy in the second approximation in SOC and do not affect the twist.

However, there is another problem with parameter $\mathcal{D}$ in Eq.~(\ref{eq:J&D}), which deals with the Keffer rule \cite{Keffer1969}. It is known that the DM interaction is of superexchange origin, i.e., the spin interaction of magnetic atoms occurs through the electrons of nonmagnetic atoms, e.g. silicon in MnSi or oxygen in Cu$_2$OSeO$_3$ \cite{GongXiang2012,Janson2014}. Besides, the DM vector $\mathbf{D}_{ij} \sim [\mathbf{r}_{\mathrm{O}i} \times \mathbf{r}_{\mathrm{O}j}]$, where $\mathbf{r}_{\mathrm{O}i}$ and $\mathbf{r}_{\mathrm{O}j}$ are the distances from the nonmagnetic atom (O) to interacting magnetic ones ($i$, $j$). Since $\mathbf{r}_{ij} = \mathbf{r}_{\mathrm{O}j} - \mathbf{r}_{\mathrm{O}i}$, the vector $\mathbf{D}_{ij}$ must be perpendicular to the bond $\mathbf{r}_{ij}$, which according to Eq.~(\ref{eq:J&D}) makes $\mathcal{D}$ equal to zero. The contradiction is partially removed by the approximate nature of the Keffer rule. A much greater effect is due to the replacement in Eq.~(\ref{eq:J&D}) of real distances $\mathbf{r}_{ij}$ with exchange distances $\tilde{\mathbf{r}}_{ij}$, which do not have to be perpendicular to $\mathbf{D}_{ij}$. However, the analysis carried out in Ref.~\cite{Chizhikov2015} for the Cu$_2$OSeO$_3$ crystal showed that the angles between the vectors $\mathbf{D}_{ij}$ and $\tilde{\mathbf{r}}_{ij}$ are close to the right angle, and thus the Keffer rule still affects the smallness of $\mathcal{D}$. So, we assume that the next order contributions to $\mathcal{D}$ can strongly affect the degree of twist. In this paper, we will show that this is indeed the case, and the key role here is played by antiferromagnetic cantings, which do not affect the twist in the second spin–orbit approximation.

\section{Discrete--continuum transition: \\ Third order approximation}
\label{sec:transition}
Let us proceed a transition from the discrete model of a cubic magnet based on the Heisenberg energy~(\ref{eq:EHeisenberg}) to the continuum approximation, following the technique developed in Ref.~\cite{Chizhikov2015}. In the latter model, the energy of a magnetic crystal is a volume integral of the GL energy density:
\begin{equation}
\label{eq:Econtinuous}
E = \int \mathcal{E}(\mathbf{r}) d\mathbf{r} ,
\end{equation}
with, from Eq.~(\ref{eq:EHeisenberg}),
\begin{equation}
\label{eq:Edensity}
\begin{split}
\mathcal{E}(\mathbf{r}) = \frac12 \left.\sum_{i,j}\right.^\prime \left( - J_{ij} \mathbf{s}_i(\mathbf{r}) \cdot \mathbf{s}_j(\mathbf{r}^\prime) + \mathbf{D}_{ij} \cdot [\mathbf{s}_i(\mathbf{r}) \times \mathbf{s}_j(\mathbf{r}^\prime)] \right) \\ - \mathbf{H} \cdot \left.\sum_i\right.^\prime g_i \mu_\mathrm{B} \mathbf{s}_i(\mathbf{r}) .
\end{split}
\end{equation}
Here, the summation is carried out over the magnetic atoms and bonds of the unit cell, and the classical spins are replaced by unimodular functions of coordinates. The number of functions coincides with the number of magnetic sublattices of the crystal. For convenience, we use the system in which the unit cell volume is equal to unity, $V_{\mathrm{u.c.}} = 1$.

Note that, in Eq.~(\ref{eq:Edensity}), the pair interactions link the function $\mathbf{s}_i$ at the point $\mathbf{r}$ with the function $\mathbf{s}_j$ at the point $\mathbf{r}^\prime = \mathbf{r} + \mathbf{r}_{ij}$, where $\mathbf{r}_{ij} \equiv \mathbf{r}_j - \mathbf{r} _i$ is the distance between the corresponding atoms in the crystal. In order to reduce all spin functions to the same argument, we use the Taylor series expansion
\begin{equation}
\label{eq:sTaylor}
\begin{split}
\mathbf{s}_j(\mathbf{r}^\prime) = [ 1 + \mathbf{r}_{ij} \cdot \boldsymbol{\nabla} + \frac12 (\mathbf{r}_{ij} \cdot \boldsymbol{\nabla})^2 + \ldots ] \mathbf{s}_j(\mathbf{r}) \\ \equiv \exp(\mathbf{r}_{ij} \cdot \boldsymbol{\nabla}) \mathbf{s}_j(\mathbf{r}) .
\end{split}
\end{equation}
Thus, spatial derivatives appear in the continuum model energy.

All terms of the GL energy can be assigned an order of smallness in terms of the spin-orbit interaction. Thus, isotropic exchange is of the zeroth order, and the DM interaction is of the first order in SOC. Since spatial modulations of magnetization in twisted magnets are due to the DM interaction, each spatial derivative can also be assigned the first order in SOC. We confine ourselves to weak magnetic fields $H < H_{c2}$, where $H_{c2} = \mathcal{D}^2 / (g \mu_\mathrm{B} \mathcal{J})$ is the field of the full unwinding of the helical magnetic structure. Thus, the magnetic field can be considered as the second order in SOC. Summarizing the above,
\begin{equation}
\label{eq:SOCorder}
D \sim \nabla \sim (v/c)^2 , \phantom{xx} g \mu_\mathrm{B} H \sim (v/c)^4 .
\end{equation}

The presence of the small parameter allows to calculate the GL energy in successive approximations. For example, in the ferromagnetic case in the zeroth approximation, all functions $\mathbf{s}_i(\mathbf{r})$ corresponding to different magnetic sublattices can be replaced by a single unimodular vector function $\boldsymbol{\mu}(\mathbf{r}) \equiv \mathbf{M}(\mathbf{r}) / |\mathbf{M}(\mathbf{r})|$ directed along the magnetization:
\begin{equation}
\label{eq:mu}
\mathbf{s}_i^{(0)}(\mathbf{r}) = \mathbf{s}_j^{(0)}(\mathbf{r}^\prime) =  \boldsymbol{\mu}(\mathbf{r}) ,
\end{equation}
with $\mathbf{M}(\mathbf{r}) = \sum_i^\prime g_i \mu_\mathrm{B} \mathbf{s}_i(\mathbf{r})$. Then it is easy to obtain the energy density in the zero approximation:
\begin{equation}
\label{eq:E0}
\mathcal{E}^{(0)} = - \frac12 \left.\sum_{i,j}\right.^\prime J_{ij} ,
\end{equation}
which is due to the isotropic exchange of collinear spins.

Now let's expand $\mathbf{s}_i(\mathbf{r})$ into components parallel and perpendicular to the magnetization:
\begin{equation}
\label{eq:siVs.ui}
\mathbf{s}_i =  \boldsymbol{\mu} \sqrt{1 - u_i^2} + \mathbf{u}_i ,
\end{equation}
where $\mathbf{u}_i \perp \boldsymbol{\mu}$ is a small spin canting, which can be represented as follows:
\begin{equation}
\label{eq:uiOrders}
\mathbf{u}_i =  \mathbf{u}_i^\prime + \mathbf{u}_i^{\prime\prime} + \mathbf{u}_i^{\prime\prime\prime} + \ldots ,
\end{equation}
with the number of primes corresponding to the approximation order. Taking into account Eqs.~(\ref{eq:siVs.ui}),~(\ref{eq:uiOrders}), successive terms in the expansion of $\mathbf{s}_i(\mathbf{r})$ can be written as
\begin{equation}
\label{eq:siOrders}
\left[\begin{array}{l}
\mathbf{s}_i^{(0)} = \boldsymbol{\mu} , \\
\mathbf{s}_i^\prime = \mathbf{u}_i^\prime , \\
\mathbf{s}_i^{\prime\prime} = \mathbf{u}_i^{\prime\prime} - \frac12 \mathbf{u}_i^{\prime2} \boldsymbol{\mu} , \\
\mathbf{s}_i^{\prime\prime\prime} = \mathbf{u}_i^{\prime\prime\prime} - (\mathbf{u}_i^\prime \cdot \mathbf{u}_i^{\prime\prime}) \boldsymbol{\mu} , \\
\ldots
\end{array}\right.
\end{equation}
A similar expansion of the function $\mathbf{s}_j(\mathbf{r}^\prime)$, taking into account Eq.~(\ref{eq:sTaylor}), contains spatial derivatives:
\begin{equation}
\label{eq:sjOrders}
\left[\begin{array}{l}
\mathbf{s}_j^{(0)} = \boldsymbol{\mu} , \\
\mathbf{s}_j^\prime = \mathbf{u}_j^\prime + (\mathbf{r}_{ij} \cdot \boldsymbol{\nabla}) \boldsymbol{\mu} , \\
\mathbf{s}_j^{\prime\prime} = \mathbf{u}_j^{\prime\prime} - \frac12 \mathbf{u}_j^{\prime2} \boldsymbol{\mu} +  (\mathbf{r}_{ij} \cdot \boldsymbol{\nabla}) \mathbf{u}_j^\prime + \frac12 (\mathbf{r}_{ij} \cdot \boldsymbol{\nabla})^2 \boldsymbol{\mu} , \\
\ldots
\end{array}\right.
\end{equation}
Here the functions of $\mathbf{r}^\prime$ on the left-hand sides of the equalities are related to the functions of $\mathbf{r}$ on the right-hand sides.

To find the canting $\mathbf{u}_i$, let's write out the part of the magnetic energy associated with the $i$th spin:
\begin{equation}
\label{eq:Ei}
E_i = - \mathbf{h}_i \cdot \mathbf{s}_i ,
\end{equation} 
where
\begin{equation}
\label{eq:hi}
\mathbf{h}_i = \sum_j \left( J_{ij} \mathbf{s}_j + [\mathbf{D}_{ij} \times \mathbf{s}_j] \right) + g_i \mu_\mathrm{B} \mathbf{H}
\end{equation}
is the local effective field affecting the spin. The equilibrium condition is that the directions of the spin and the local field coincide:
\begin{equation}
\label{eq:equilibrium}
\mathbf{s}_i = \mathbf{h}_i / |\mathbf{h}_i| .
\end{equation}
Comparison of Eqs.~(\ref{eq:siVs.ui}) and (\ref{eq:equilibrium}) gives
\begin{equation}
\label{eq:uiVs.hi}
\mathbf{u}_i = (\mathbf{h}_i - \boldsymbol{\mu} (\boldsymbol{\mu} \cdot \mathbf{h}_i)) / |\mathbf{h}_i| .
\end{equation} 

The perturbation theory is also applicable to the effective field:
\begin{equation}
\label{eq:hiOrders}
\mathbf{h}_i = \mathbf{h}_i^{(0)} + \mathbf{h}_i^{\prime} + \mathbf{h}_i^{\prime\prime} + \ldots
\end{equation} 
For example, the effective fields in the zeroth and first approximations in SOC are
\begin{equation}
\label{eq:hi0}
\mathbf{h}_i^{(0)} = \sum_j J_{ij} \boldsymbol{\mu}
\end{equation}
and
\begin{equation}
\label{eq:hi1}
\mathbf{h}_i^{\prime} = \sum_j \left(J_{ij} (\mathbf{r}_{ij} \cdot \boldsymbol{\nabla}) \boldsymbol{\mu} + J_{ij} \mathbf{u}_j^{\prime} + [\mathbf{D}_{ij} \times \boldsymbol{\mu}]\right) ,
\end{equation} 
correspondingly. Note that due to the definition of cantings and the unimodularity of $\boldsymbol{\mu}(\mathbf{r})$, $\mathbf{h}_i^{\prime} \cdot \boldsymbol{\mu} = 0$. Then the first-order canting is
\begin{equation}
\label{eq:ui1}
\mathbf{u}_i^{\prime} = \frac{\mathbf{h}_i^{\prime}}{|\mathbf{h}_i^{(0)}|} = \frac{\sum_j \left(J_{ij} (\mathbf{r}_{ij} \cdot \boldsymbol{\nabla}) \boldsymbol{\mu} + J_{ij} \mathbf{u}_j^{\prime} + [\mathbf{D}_{ij} \times \boldsymbol{\mu}]\right)}{\sum_j J_{ij}} .
\end{equation}

As mentioned above, since the Heisenberg energy (\ref{eq:EHeisenberg}) does not explicitly contain atomic coordinates, they cannot affect the coefficients of the GL free energy (\ref{eq:ELandau}). This means that the cantings $\mathbf{u}^{\prime}$ must include something to compensate for the phase shifts between magnetic sublattices. We can take this into account by introducing fictitious ``exchange'' coordinates defined by the system of linear equations
\begin{equation}
\label{eq:rex}
\sum_j J_{ij} (\mathbf{r}_{j,\mathrm{ex}} - \mathbf{r}_{i,\mathrm{ex}}) = 0 .
\end{equation}
When solving the system (\ref{eq:rex}), it is necessary to take into account the symmetry of the crystal. For example, in cubic magnets with the $B20$ structure (MnSi, etc.), all magnetic atoms are in same Wyckoff positions $4a$ $(x, x, x)$ of the space group $P2_13$ \cite{IntTables}, and the system reduces to a single linear equation that determines the exchange coordinate $x_\mathrm{ex}$ \cite{Chizhikov2013}. In what follows, the Taylor expansions will include exchange coordinates instead of real ones, i.e. $\mathbf{s}_j(\mathbf{r}^\prime) = \exp(\tilde{\mathbf{r}}_{ij} \cdot \boldsymbol{\nabla}) \mathbf{s}_j(\mathbf{r})$, where $\tilde{\mathbf{r}}_{ij} \equiv \mathbf{r}_{ij,\mathrm{ex}} = \mathbf{r}_{j,\mathrm{ex}} - \mathbf{r}_{i,\mathrm{ex}}$. When taking into account the exchange coordinates defined in this way, the first term in the right side numerator of Eq.~(\ref{eq:ui1}) is equal to zero, and this equation can be rewritten as
\begin{equation}
\label{eq:ui1-equation}
\sum_j J_{ij} (\mathbf{u}_i^{\prime} - \mathbf{u}_j^{\prime}) = \sum_j [\mathbf{D}_{ij} \times \boldsymbol{\mu}] .
\end{equation}
A solution can be found in the form
\begin{equation}
\label{eq:ui1Vs.rhoi}
\mathbf{u}_i^{\prime} = [\boldsymbol{\rho}_i \times \boldsymbol{\mu}] ,
\end{equation}
where associated with magnetic atoms vectors $\boldsymbol{\rho}_i$ can be calculated from the system
\begin{equation}
\label{eq:rho}
\sum_j J_{ij} (\boldsymbol{\rho}_i - \boldsymbol{\rho}_j) = \sum_j \mathbf{D}_{ij} ,
\end{equation}
provided that they satisfy the symmetry of the crystal.

The first-order cantings $\mathbf{u}_i^{\prime}$ are sufficient to calculate the GL free energy in the third approximation in SOC (see Appendix).

Thus, the first-order contribution $\mathcal{E}^{(1)}$ to the energy density is zero, and the second-order contribution is
\begin{equation}
\label{eq:E2}
\mathcal{E}^{(2)} = \frac12 \mathcal{J} \frac{\partial \boldsymbol{\mu}}{\partial r_\alpha} \cdot \frac{\partial \boldsymbol{\mu}}{\partial r_\alpha} + \mathcal{D}_1 \boldsymbol{\mu} \cdot \mathrm{curl} \boldsymbol{\mu} - \mathbf{H} \cdot (M_0 \boldsymbol{\mu}) + C_2 ,
\end{equation}
with coefficients $\mathcal{J}$ and $\mathcal{D}_1$ being calculated by Eq.~(\ref{eq:J&D}), where the physical distances $\mathbf{r}_{ij}$ are replaced with the exchange distances $\tilde{\mathbf{r}}_{ij}$:
\begin{equation}
\label{eq:Jcal}
\mathcal{J} = \frac16 \left.\sum_{i,j}\right.^\prime J_{ij} \tilde{r}_{ij}^2 , 
\end{equation}
\begin{equation}
\label{eq:Dcal1}
\mathcal{D}_1 = - \frac16 \left.\sum_{i,j}\right.^\prime \mathbf{D}_{ij} \cdot \tilde{\mathbf{r}}_{ij} ,
\end{equation}
and the constant term is
\begin{equation}
\label{eq:C2}
C_2 = - \frac16 \left.\sum_{i,j}\right.^\prime J_{ij} (\boldsymbol{\rho}_i - \boldsymbol{\rho}_j)^2 .
\end{equation}

Finally, the third-order energy density is
\begin{equation}
\label{eq:E3}
\mathcal{E}^{(3)} = \mathcal{D}_2 \boldsymbol{\mu} \cdot \mathrm{curl} \boldsymbol{\mu} + C_3 ,
\end{equation}
with
\begin{equation}
\label{eq:Dcal2}
\begin{split}
\mathcal{D}_2 = & \frac1{12} \left.\sum_{i,j}\right.^\prime J_{ij} \tilde{\mathbf{r}}_{ij} \cdot [\boldsymbol{\rho}_i \times \boldsymbol{\rho}_j] \\ & + \frac1{12} \left.\sum_{i,j}\right.^\prime \mathbf{D}_{ij} \cdot [\tilde{\mathbf{r}}_{ij} \times (\boldsymbol{\rho}_i + \boldsymbol{\rho}_j)] ,
\end{split}
\end{equation}
\begin{equation}
\label{eq:C3}
C_3 = \frac16 \left.\sum_{i,j}\right.^\prime \mathbf{D}_{ij} \cdot [\boldsymbol{\rho}_i \times \boldsymbol{\rho}_j] .
\end{equation}
As we can see, the second-order correction to the DM parameter $\mathcal{D}$ is fully determined by the antiferromagnetic cantings $\mathbf{u}_i^\prime$ related to the rotation vectors $\boldsymbol{\rho}_i$. In the next section, using the cubic ferrimagnet Cu$_2$OSeO$_3$ as an example, we will show how this correction affects the pitch of the magnetic helices.

\section{Helimagnetic $\mathbf{Cu_2OSeO_3}$ }
\label{sec:example}
The formulas of the previous section refer to the case of a ferromagnetic crystal, in which neighboring spins are almost codirectional, and the magnetic structure varies slowly on scales much larger than the unit cell. The Cu$_2$OSeO$_3$ crystal belongs to a different type of magnets, namely, it is a collinear ferrimagnet. Like the $B20$-type structures, the crystal is described by the space group $P2_13$, and its unit cell contains sixteen magnetic copper atoms in two different Wyckoff positions \cite{Effenberger1986}:
\begin{equation}
\label{eq:CuPositions}
\begin{array}{ll}
4a & (x_1, x_1, x_1) = (0.8860, 0.8860, 0.8860) , \\
12b & (x_2, y_2, z_2) = (0.1335, 0.1211, 0.8719) .
\end{array}
\end{equation}
The magnetic moments of all copper atoms are the same in magnitude, but not in direction. Four spins in the special position $4a$, lying on the three-fold axes of the cubic crystal, are approximately opposite to twelve spins in the general position $12b$. Thus, the total magnetic moment of the unit cell is two times less compared to the ferromagnetic ordering and is codirectional with the spins in the $12b$ position. Due to the absence of an inversion center, the local ferrimagnetic structure rotates on scales much larger than the unit cell, which leads to the appearance of either a helical phase or a rather unusual skyrmionic phase with double twist of the magnetization field \cite{SekiSci2012,AdamsPRL2012,SekiPRB2012,OnosePRL2012,SekiPRB2012-2,White2012}. The possibility of observing even more intriguing magnetic structures, such as coupled magnetic monopoles \cite{Mueller2020} and tilted skyrmion and spiral states \cite{Leonov2023}, is also discussed.

\begin{figure}[h]
	\begin{center}
		\includegraphics[width=7cm]{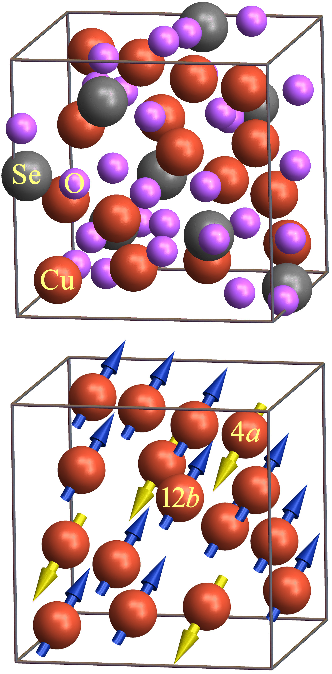}
		\caption{\label{fig:unitcell} (Color online) The unit cell of the cubic ferrimagnet Cu$_2$OSeO$_3$, containing eight formula units, and its magnetic sublattice. The crystal symmetry is described by the space group $P2_13$. Below the Curie point $T_\mathrm{C} = 58$ K, the spins of copper atoms are almost collinear and have the opposite directions in the Wyckoff positions $4a$ and $12b$.}
	\end{center}
\end{figure}

Using a simple mathematical trick, the formulas derived for a ferromagnetic crystal can be used without modification for the case of collinear ferrimagnetic. First, we change the sign of the constant $g_i$ for all magnetic moments in the $4a$ position. Second, for all bonds connecting atoms in different Wyckoff positions, we change the signs of the isotropic exchange constants $J_{ij}$ and the DM vectors $\mathbf{D}_{ij}$. This allows us to assume that all classical spins in the unit cell are directed along magnetization, while the real magnetic moments can be opposite.

In Refs.~\cite{GongXiang2012,Janson2014}, the DFT calculations of the exchange parameters of bonds in Cu$_2$OSeO$_3$ were carried out. Using these data, we calculated the continuum model constants $\mathcal{J}$ and $\mathcal{D}$ in the second approximation in SOC \cite{Chizhikov2015}. Let us briefly describe the results obtained, referring to Ref.~\cite{Chizhikov2015} for details.

The initial data are the isotropic exchange constants $J_1$--$J_5$ and the DM vectors $\mathbf{D}_1$--$\mathbf{D}_5$ calculated in Ref.~\cite{GongXiang2012} for five nonequivalent bonds (Table~\ref{tab:JD}). Two nonequivalent ferromagnetic bonds ($J > 0$) connect atoms in the position $12b$. Three other bonds are antiferromagnetic ($J < 0$) and they connect atoms in different Wyckoff positions. Note that an atom can have more than one bond of each type. Thus, each atom in the Wyckoff position $12b$ has four neighbors in the same position and three neighbors in the $4a$ position. Each atom in the position $4a$ has nine neighbors in the position $12b$. The DM vectors are approximately satisfy the Keffer rule: the angle between the DM vector $\mathbf{D}_{ij}$ and the bond $\mathbf{r}_{ij}$ varies in the range 85.3--99.5$^\circ$, and the average deviation from the right angle is 5.3$^\circ$.

\begin{table*}
	\caption{\label{tab:JD} Nonequivalent bonds of the copper atom in the position $(x_2, y_2, z_2)$ with its magnetic neighbors and the exchange parameters \cite{GongXiang2012}.}
	\begin{ruledtabular}
	\begin{tabular}{cccdcc}
		n & neighboring Cu atom &  Wyckoff pos. & \multicolumn{1}{r}{$J$, meV} & $\mathbf{D}$, meV & $\angle (\mathbf{D}_{ij}, \mathbf{r}_{ij})$ \\
		\hline
		1 & $(z_2 - \frac12, \frac12 - x_2, 1 - y_2)$ & $12b$ & 1.132 & $(0.289, -0.325, -0.051)$ & 94.3$^\circ$ \\
		2 & $(x_1 - 1, x_1 - 1, x_1)$ & $4a$ & -6.534 & $(1.120, -1.376, 0.300)$ & 85.3$^\circ$ \\
		3 & $(z_2 - 1, x_2, 1 + y_2)$ & $12b$ & 3.693 & $(-0.263, 0.167, -0.407)$ & 99.5$^\circ$ \\
		4 & $(1 - x_1, x_1 - \frac12, \frac32 - x_1)$ & $4a$ & -0.900 & $(-0.490, 1.238, 1.144)$ & 86.3$^\circ$ \\
		5 & $(\frac12 - x_1, 1 - x_1, x_1 - \frac12)$ & $4a$ & -0.984 & $(0.045, -0.087, -0.059)$ & 85.8$^\circ$ \\
	\end{tabular}
	\end{ruledtabular}
\end{table*}

The first step is to calculate the exchange coordinates of the magnetic atoms. Using the symmetry of the crystal, Eq.~(\ref{eq:rex}) reduces to a system of four linear equations for the unknowns $x_{1,\mathrm{ex}}$, $x_{2,\mathrm{ex}}$, $y_ {2,\mathrm{ex}}$, $z_{2,\mathrm{ex}}$. All the coefficients and constant terms in the equations are linear combinations of the isotropic exchange constants $J_1$--$J_5$. Consequently, the exchange coordinates have the form $P_4(J_1, \ldots J_5) / Q_4(J_1, \ldots J_5)$, where $P_4$ and $Q_4$ are fourth-degree polynomials in the exchange constants. The result of routine calculations with the data from Table~\ref{tab:JD} is
\begin{equation}
\label{eq:Cu-rex}
\begin{array}{l}
x_{1,\mathrm{ex}} = 0.942 , \\
x_{2,\mathrm{ex}} = -0.004 , \phantom{x} y_{2,\mathrm{ex}} = 0.020 , \phantom{x} z_{2,\mathrm{ex}} = 0.897 .
\end{array}
\end{equation}
Then, using Eqs.~(\ref{eq:Jcal}),~(\ref{eq:Dcal1}), we calculate the exchange parameters of the continuum model in the second approximation in SOC: $\mathcal{J} = 5.130$ meV, $\mathcal{D}_1 = 0.970$ meV. Without taking into account corrections of the next orders, the wavenumber of magnetic helicoids is $k = \mathcal{D}_1 / \mathcal{J} = 0.189$, and the pitch is $p = 2 \pi / |k| = 33.2$ in the parameters of the cubic lattice. The experimental values known from small-angle neutron scattering data are $k_\mathrm{exp} = 0.088$ and $p_\mathrm{exp} = 71.4$, respectively. Note that the angles between the DM vectors $\mathbf{D}_{ij}$ and the exchange distances $\tilde{\mathbf{r}}_{ij}$ are less consistent with the Keffer rule: they vary in the range 71.3--110.2$^\circ$ with the average deviation from the right angle of 11.5$^\circ$.

In order to find the correction $\mathcal{D}_2$ according to Eq.~(\ref{eq:Dcal2}), it is necessary to calculate the vectors $\boldsymbol{\rho}_1$ and $\boldsymbol{\rho}_2$, which determine the spin cantings of the atoms $(x_1, x_1, x_1)$ and $(x_2, y_2, z_2)$, respectively. As in the case of exchange coordinates, the symmetry reduces Eq.~(\ref{eq:rho}) to a system of four linear equations with the unknowns $\rho_{1x}$, $\rho_{2x}$, $\rho_{2y}$, $\rho_{2z}$. In addition, all the coefficients of the equations are linear combinations of the exchange constants $J_1$--$J_5$, and the constant terms are combinations of components of the vectors $\mathbf{D}_1$--$\mathbf{D}_5$. This means that the components of $\boldsymbol{\rho}_1$, $\boldsymbol{\rho}_2$ have the form $P_{3,1}(J_1, \ldots J_5; \mathbf{D}_1, \ldots \mathbf{D}_1) / Q_4(J_1, \ldots J_5)$, where $P_{3,1}$ are polynomials linear in the components of the vectors $\mathbf{D}_1$--$\mathbf{D}_5$ and third-degree in $J_1$--$J_5$. Routine calculation with the data from Table~\ref{tab:JD} gives
\begin{equation}
\label{eq:Cu-rho}
\begin{array}{l}
\boldsymbol{\rho}_1 = (-0.049, -0.049, -0.049) , \\
\boldsymbol{\rho}_2 = (-0.123, -0.034, -0.159) .
\end{array}
\end{equation}
Note that the vector $\boldsymbol{\rho}_1$ corresponds to the canting of the classical spin, which according to our trick is opposite to the real spin of the atom. The cantings of the real magnetic moments in the $4a$ position have the opposite sign.

The calculation of the second-order spin-orbit correction to the DM constant by Eq.~(\ref{eq:Dcal2}) gives $\mathcal{D}_2 = -0.369$ meV. In total, $\mathcal{D} = \mathcal{D}_1 + \mathcal{D}_2 = 0.600$ meV, the helicoid wavenumber is $k = \mathcal{D} / \mathcal{J} = 0.117$, and the helix pitch is $p = 53.7$ in lattice parameters. The positive value of $k$ means that in the considered enantiomorph of the Cu$_2$OSeO$_3$ crystal only right-handed helicoids can appear. It can be seen that taking into account the third-order terms in the GL energy leads to a change in the wavenumber of helices by more than one and a half times. This means that the antiferromagnetic cantings, whose influence on twist was previously neglected, can significantly affect the mesoscopic magnetic structure of helimagnets. For example, in the case of Cu$_2$OSeO$_3$, a significant weakening of helicity occurs.

We should note the fact that the calculated value of the wavenumber, $k_\mathrm{calc} = 0.117$, still differs significantly from the experimental value $k_\mathrm{exp} = 0.088$. In order to understand whether the discrepancy is due to the limitation to the third order in SOC or to the insufficient accuracy of the {\it ab initio} calculation of the exchange constants of bonds, a spin helicoid was simulated using the same constants from Table~\ref{tab:JD}. The simulation was carried out for periodic spin helicoids with pitches $(n, 0, 0)$, $(n, n, 0)$, and $(n, n, n)$ directed along crystallographic directions $[100]$, $[110]$, and $[111]$, respectively. The integer parameter $n$ for each of the directions went through all the values within certain limits, so that the wavenumber of helicoids varied within 0.06--0.17. For every pitch, the elongated crystal cell was filled with one turn of a roughly made helicoid, after which the spin structure relaxed using Eqs.~(\ref{eq:hi}),~(\ref{eq:equilibrium}).

\begin{figure}[h]
	\begin{center}
		\includegraphics[width=7cm]{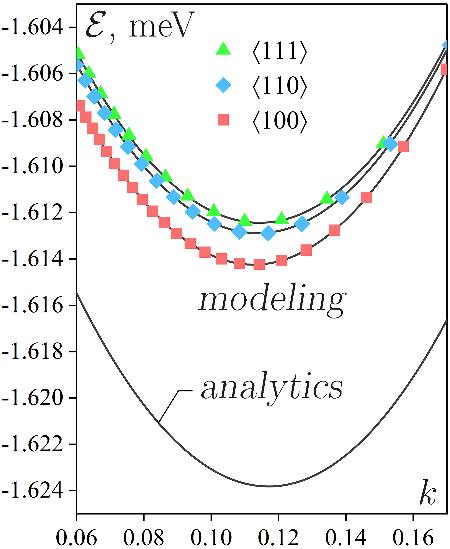}
		\caption{\label{fig:parabolae} (Color online) Calculated dependence of the average magnetic energy density of the Cu$_2$OSeO$_3$ ferrimagnet on the helicoid wavenumber $k = 2 \pi / p$: comparison of the analytical expression~(\ref{eq:parabola}) and the numerical simulation results. In both cases, the exchange constants from Table~\ref{tab:JD} are used, and the energy is measured from the level $\mathcal{E}^{(0)} = -158.916$ meV corresponding to the exactly collinear spins. The helicoid pitch is measured in the unit cell parameters.}
	\end{center}
\end{figure}

Fig.~\ref{fig:parabolae} shows the calculated dependences of the helicoid energy per unit cell of the crystal on the wavenumber $k$. The energy is measured from the level $\mathcal{E}^{(0)} = -158.916$ meV calculated for collinear spins by Eq.~(\ref{eq:E0}) with the exchange constants $J_1$--$J_5$ from Table~\ref{tab:JD}. The results of modeling helicoids for various crystallographic directions are well approximated by parabolas. For comparison, the analytical curve
\begin{equation}
\label{eq:parabola}
\mathcal{E} (k) = \frac12 \mathcal{J} k^2 - \mathcal{D} k + C_2 + C_3
\end{equation}
is shown, which corresponds to the GL energy density calculated using Eqs.~(\ref{eq:E2}),~(\ref{eq:E3}) for the helicoidally twisted field $\boldsymbol{\mu} = (\cos k z, \sin k z, 0)$. The constants calculated by Eqs.~(\ref{eq:C2}), (\ref{eq:C3}) are $C_2 = -1.573$ meV, $C_3 = -0.016$ meV.

It is evident from Fig.~\ref{fig:parabolae} that the plot points for every nonequivalent crystallographic direction fit their own smooth curve. This is a consequence of the cubic anisotropy of the crystal, which manifests itself starting from the fourth-order contributions to the energy density. Taking into account the anisotropic terms, the dependences $\mathcal{E}(k)$ for helicoids should actually be described by the fourth-degree polynomials in $k$. Finding the coefficients of these polynomials is, in fact, an ill-posed problem, the solution of which strongly depends both on the accuracy of calculating the discrete points of the graphs and on the fitting method. However, it is easy to show that in the fourth approximation in SOC, the anisotropy is introduced by the combination $n_x^4 + n_y^4 + n_z^4$, where $\mathbf{n}$ is the helix axis direction. For example, for the crystallographic directions $\left\langle 100 \right\rangle$, $\left\langle 110 \right\rangle$, and $\left\langle 111 \right\rangle$ this combination is equal to $1$, $1/2$, and $1/3$, respectively. Then we can restrict ourselves to the parabolic dependence of the helicoid energy on the wavenumber: 
\begin{equation}
\label{eq:parabola2}
\begin{split}
\mathcal{E}(k) = & \alpha_2 k^2 + \alpha_1 k + \alpha_0 \\ & + (\beta_2 k^2 + \beta_1 k + \beta_0) (n_x^4 + n_y^4 + n_z^4) .
\end{split}
\end{equation}
Fitting the discrete plots with Eq.~(\ref{eq:parabola2}) using the least squares method gives the following values of the coefficients: $\alpha_0 = -1.579$ meV, $\alpha_1 = -0.566$ meV, $\alpha_2 = 2.457$ meV, $\beta_0 = -0.003 $ meV, $\beta_1 = -0.012$ meV, $\beta_2 = 0.113$ meV. The stiffness of the magnetic structure and the DM constant now depend on the direction $\mathbf{n}$:
\begin{equation}
\label{eq:J&Danisotropy}
\begin{array}{l}
\mathcal{J} = 2 \alpha_2 + 2 \beta_2 (n_x^4 + n_y^4 + n_z^4) , \\
\mathcal{D} = - \alpha_1 - \beta_1 (n_x^4 + n_y^4 + n_z^4) .
\end{array}
\end{equation}
The parameters calculated by Eq.~(\ref{eq:J&Danisotropy}) are given in Table~\ref{tab:JDanisotropy}, and the approximation accuracy is obvious from the fitting parabolas in Fig.~\ref{fig:parabolae}.

\begin{table}
	\caption{\label{tab:JDanisotropy} Results of modeling spin helicoids with the axes parallel to the crystallographic directions $\left\langle 100 \right\rangle$, $\left\langle 110 \right\rangle$, $\left\langle 111 \right\rangle$. The simulation was performed with the exchange constants from Table~\ref{tab:JD}.}
	\begin{ruledtabular}
	\begin{tabular}{cddd}
		& \multicolumn{1}{r}{$\mathcal{J}$, meV} & \multicolumn{1}{r}{$\mathcal{D}$, meV} & \multicolumn{1}{r}{$k = \mathcal{D} / \mathcal{J}$} \\
		\hline
		modeling: &  &  &  \\
		$\left\langle 100 \right\rangle$ & 5.1405 & 0.5788 & 0.1126  \\
		$\left\langle 110 \right\rangle$ & 5.0274 & 0.5727 & 0.1139 \\
		$\left\langle 111 \right\rangle$ & 4.9887 & 0.5705 & 0.1143  \\
		analytics & 5.130 & 0.600 & 0.117 \\
		experiment &  &  & 0.088 \\
	\end{tabular}
	\end{ruledtabular}
\end{table}

Comparison of the simulation results with the values of the exchange constants calculated by the analytical formulas (\ref{eq:Jcal}), (\ref{eq:Dcal1}) and (\ref{eq:Dcal2}) shows minor difference within a few percent, and, therefore, the difference from the experimental data is due to the initial parameters used in the calculations (Table~\ref{tab:JD}). The seemingly significant difference between the minima of the parabola~(\ref{eq:parabola}) and the simulation curves is due to the unaccounted for constant terms of the fourth order in SOC, some of which are comparable to $C_3$. For example,
\begin{equation}
\label{eq:C4-1}
\frac1{16} \left.\sum_{i,j}\right.^\prime J_{ij} (\mathbf{u}_i^{\prime2} - \mathbf{u}_j^{\prime2})^2 = 0.0103 \pm 0.0003 \phantom{x} \mathrm{meV} ,
\end{equation}
\begin{equation}
\label{eq:C4-2}
\begin{split}
- \frac14 \left.\sum_{i,j}\right.^\prime \mathbf{D}_{ij} \cdot (\mathbf{u}_i^{\prime2} [\boldsymbol{\mu} \times \mathbf{u}_j^\prime] + \mathbf{u}_j^{\prime2} [\mathbf{u}_i^\prime \times \boldsymbol{\mu}]) \\ = 0.0127 \pm 0.0008 \phantom{x} \mathrm{meV} ,
\end{split}
\end{equation}
where a small uncertainty is caused by the cubic anisotropy. In addition, the energy $\mathcal{E}^{(4)}$ must include terms related to the second-order cantings $\mathbf{u}_i^{\prime\prime}$, which are not considered in the paper.

\section{Discussion}
\label{sec:discussion}
Summarizing the above, we found that antiferromagnetic spin cantings, which are usually imperceptible due to their smallness, have a significant effect on the mesoscopic structure of helimagnets, changing the wavenumber $k$ of magnetic helicoids. In the analyzed example of the Cu$_2$OSeO$_3$ ferrimagnet, the influence of the cantings leads to an increase in the helix pitch, but it can also be vice versa. Indeed, suppose that all DM vectors $\mathbf{D}_{ij}$ in Table~\ref{tab:JD} change their signs. According to Eq.~(\ref{eq:rho}), this would mean that the rotation vectors $\boldsymbol{\rho}_i$ also change signs. It is evident from Eqs.~(\ref{eq:Dcal1}),~(\ref{eq:Dcal2}) that the first-order DM parameter $\mathcal{D}_1$ changes its sign, while the second-order correction $\mathcal{D}_2$ does not. Consequently, the modulus of the wavenumber $|k|$ decreases, and the helix pitch $p = 2 \pi / |k|$ increases. Note that in this case the wavenumber is negative, which corresponds to the left-handed helicoid. However, it should be distinguished from the case of spatial inversion, which also leads to a change in the handedness. Indeed, upon inversion, the crystal transforms into its mirror enantiomorph, so the signs of the interatomic distances $\tilde{\mathbf{r}}_{ij}$ change, but the pseudovectors $\mathbf{D}_{ij} $ and $\boldsymbol{ \rho}_i$ retain their signs. Then, according to Eqs.~(\ref{eq:Dcal1}),~(\ref{eq:Dcal2}), the sign of the wavenumber changes, but the modulus is preserved.

\begin{figure}[h]
	\begin{center}
		\includegraphics[width=7cm]{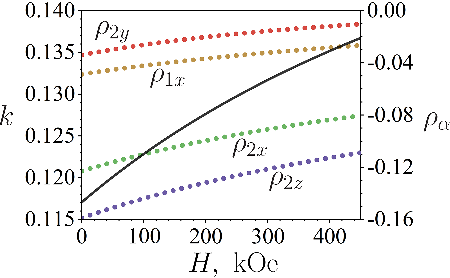}
		\caption{\label{fig:k(H)} (Color online) Calculated dependencies of the wavenumber $k = \mathcal{D} / \mathcal{J}$ of magnetic helicoids and the components of the $\boldsymbol{\rho}_i$ vectors on the external magnetic field $H$ for the Cu$_2$OSeO$_3$ crystal. The spin exchange constants from Table~\ref{tab:JD} are used. The dependencies for $\rho_{i\alpha}$ start from the values specified in Eq.~(\ref{eq:Cu-rho}) at $H = 0$.}
	\end{center}
\end{figure}

The correction $\mathcal{D}_2$ to the DM parameter depends on the magnitude of spin cantings. In turn, the cantings can vary under the action of a strong external field $H \gg H_{c2}$, where $H_{c2}$ is the field of complete unwinding of spin helices. When the field becomes comparable in magnitude to the exchange interaction, its contribution to the effective field $\mathbf{h}_i^{(0)}$ can no longer be neglected. Since in the polarized phase $\boldsymbol{\mu} \parallel \mathbf{H}$, Eq.~(\ref{eq:hi0}) should be rewritten as
\begin{equation}
\label{eq:hi0-2}
\mathbf{h}_i^{(0)} = \left( \sum_j J_{ij} + g_i \mu_\mathrm{B} H \right) \boldsymbol{\mu} ,
\end{equation} 
and the system~(\ref{eq:rho}) for rotation vectors now looks like
\begin{equation}
\label{eq:rho-2}
g_i \mu_\mathrm{B} H \boldsymbol{\rho}_i + \sum_j J_{ij} (\boldsymbol{\rho}_i - \boldsymbol{\rho}_j) = \sum_j \mathbf{D}_{ij} .
\end{equation}m
Fig.~\ref{fig:k(H)} shows the dependences of the components of the $\boldsymbol{\rho}_i$ vectors and the wavenumber $k$ on the external magnetic field up to 450~kOe, calculated for Cu$_2$OSeO$_3$ with the exchange constants from Table~\ref{tab:JD}. Note that for this crystal $H_{c2} \sim 1$~kOe \cite{AdamsPRL2012}. Although there is no helical order if $H > H_{c2}$, the wavenumber $k$ still can be measured with magnons. In the polarized phase, the DM interaction still affects the dispersion relation of spin waves, leading to a shift of the magnon spectrum by the value of $k$ for magnons propagating along the field \cite{Kataoka1987}. It is worth noting that although the magnon spectrum of the Cu$_2$OSeO$_3$ crystal has been studied repeatedly \cite{Laurita2017,Grigoriev2019,YiLuo2020,Zhang2020}, no one has consciously studied the change in the wavenumber in a strong magnetic field.

The numerical simulation results shown in Fig.~\ref{fig:parabolae} demonstrate anisotropy well described by the cubic invariant $n_x^4 + n_y^4 + n_z^4$, where $\mathbf{n}$ is the direction of the helicoid axis. Note that this anisotropy arises exclusively due to the DM interaction. Moreover, the antiferromagnetic cantings again play a significant role here. For example, coefficient $\beta_0$ in Eq.~(\ref{eq:parabola2}), which makes a constant anisotropic contribution to the energy $\mathcal{E}(k)$, is related to the fourth-order terms~(\ref{eq:C4-1}) and (\ref{eq:C4-2}), which explicitly contain the cantings $\mathbf{u}^\prime_i$ (the second-order cantings $\mathbf{u}^{\prime\prime}_i$ are not considered here). Another possible mechanism for the appearance of anisotropy is the interaction of spins with a local crystal field. In the latter case, the energy does not contain anisotropic terms with spatial derivatives and, accordingly, $\beta_1 = \beta_2 = 0$. It was shown in Ref.~\cite{Chizhikov2021} that this mechanism contradicts the experimental data on the anisotropy of the $A$ phase stability in MnSi and Cu$_2$OSeO$_3$.

\section*{Acknowledgements}
We are grateful to S.~V.~Grigoriev, S.~Grytsiuk, I.~V.~Kashin, S.~N.~Andreev, and V.~V.~Mazurenko for fruitful discussions. This work was supported by the Ministry of Science and Higher Education within the State assignment of Federal Scientific Research Center ``Crystallography and Photonics'' of Russian Academy of Sciences.

\appendix 
\section{}
Let us pass from the energy density~(\ref{eq:Edensity}), which contains spin fields related to different magnetic sublattices, to the GL energy density~(\ref{eq:ELandau}), which depends on the single vector field $\boldsymbol{\mu}(\mathbf{r})$. To do this, we will sequentially collect together the terms of the same order in SOC, taking into account the expansions~(\ref{eq:siOrders}),~(\ref{eq:sjOrders}). For example, the first-order contribution to the energy is
\begin{equation}
\label{eq:E1}
\begin{split}
\mathcal{E}^{(1)} = \frac12 \left.\sum_{i,j}\right.^\prime \{ - J_{ij} \boldsymbol{\mu} \cdot (\tilde{\mathbf{r}}_{ij} \cdot \boldsymbol{\nabla}) \boldsymbol{\mu} - J_{ij} (\mathbf{u}_i^\prime \cdot \boldsymbol{\mu} + \boldsymbol{\mu} \cdot \mathbf{u}_j^\prime) \\ + \mathbf{D}_{ij} [\boldsymbol{\mu} \times \boldsymbol{\mu}] \} = 0 ,
\end{split}
\end{equation}
where the first term in curly brackets is equal to zero due to the unimodularity of the field $\boldsymbol{\mu}(\mathbf{r})$ ($\boldsymbol{\nabla} \boldsymbol{\mu}^2 = 0$), whereas the second one is zero due to the definition of the spin cantings ($\mathbf{u}_i \perp \boldsymbol{\mu}$).

Similarly, it is easy to show that the cantings $\mathbf{u}_i^{\prime\prime}$ and $\mathbf{u}_i^{\prime\prime\prime}$ do not contribute to $\mathcal{E}^{(2)}$ and $\mathcal{E}^{(3)}$, correspondingly:
\begin{equation}
\label{eq:upp2E2}
\begin{split}
- \frac12 \left.\sum_{i,j}\right.^\prime J_{ij} (\mathbf{u}_i^{\prime\prime} \cdot \boldsymbol{\mu} + \boldsymbol{\mu} \cdot \mathbf{u}_j^{\prime\prime}) = 0 , \\
- \frac12 \left.\sum_{i,j}\right.^\prime J_{ij} (\mathbf{u}_i^{\prime\prime\prime} \cdot \boldsymbol{\mu} + \boldsymbol{\mu} \cdot \mathbf{u}_j^{\prime\prime\prime}) = 0 .
\end{split}
\end{equation}
Even more surprising is that the second-order cantings $\mathbf{u}_i^{\prime\prime}$ also do not contribute to the third-order energy density $\mathcal{E}^{(3)}$. For example, the contribution
\begin{equation}
\label{eq:upp2E3-1}
\begin{split}
- \frac12 & \left.\sum_{i,j}\right.^\prime J_{ij} \mathbf{u}_i^{\prime\prime} \cdot (\tilde{\mathbf{r}}_{ij} \cdot \boldsymbol{\nabla}) \boldsymbol{\mu} \\ & = - \frac12 \left.\sum_i\right.^\prime \mathbf{u}_i^{\prime\prime} \cdot \left( \sum_j J_{ij} \tilde{\mathbf{r}}_{ij} \cdot \boldsymbol{\nabla} \right) \boldsymbol{\mu} = 0
\end{split}
\end{equation}
is zero due to the definition of the exchange coordinates, see Eq.~(\ref{eq:rex}). Let's combine the rest of the third-order terms with $\mathbf{u}_i^{\prime\prime}$:
\begin{equation}
\label{eq:upp2E3-2}
\begin{split}
\frac12 \left.\sum_{i,j}\right.^\prime \left\{ - J_{ij} \mathbf{u}_i^{\prime\prime} \cdot \mathbf{u}_j^\prime + J_{ij} ( \mathbf{u}_i^\prime \cdot \mathbf{u}_i^{\prime\prime}) \boldsymbol{\mu} \cdot \boldsymbol{\mu} + \mathbf{D}_{ij} \cdot [\mathbf{u}_i^{\prime\prime} \times \boldsymbol{\mu}] \right\} \\
= \frac12 \left.\sum_i\right.^\prime \mathbf{u}_i^{\prime\prime} \cdot \sum_j \left\{ J_{ij} (\mathbf{u}_i^{\prime} - \mathbf{u}_j^{\prime}) - [\mathbf{D}_{ij} \times \boldsymbol{\mu}] \right\} = 0 .
\end{split}
\end{equation}
Here the equality to zero is due to Eq.~(\ref{eq:ui1-equation}) for the first-order cantings. The terms containing $\mathbf{u}_j^{\prime\prime}$ disappear in the same way. It is used that the calculated sums are taken over all Cu--Cu bonds of the unit cell and are symmetric in the permutation of the summation indices $i \leftrightarrow j$, subject to the obvious relations: $J_{ji} = J_{ij}$, $\mathbf{D}_{ji} = - \mathbf{D}_{ij}$, $\tilde{\mathbf{r}}_{ji} = - \tilde{\mathbf{r}}_{ij}$. Thus, the cantings of the second and third orders, $\mathbf{u}_i^{\prime\prime}$ and $\mathbf{u}_i^{\prime\prime\prime}$, do not contribute to the energy up to the third approximation in SOC inclusive.

Above, to calculate some (zero) contributions to the energy, we used (i) the unimodularity of the vector field $\boldsymbol{\mu}(\mathbf{r})$, (ii) the perpendicularity of the antiferromagnetic cantings $\mathbf{u}_i$ to the magnetization, (iii) the symmetry under permutation of summation indices $i,j$. In addition, one can also use the fact that the summation over all bonds is similar to averaging over the point group of the crystal. Indeed, the vectors $\tilde{\mathbf{r}}_{ij}$, $\mathbf{D}_{ij}$ associated with equivalent interatomic bonds are related to each other by symmetry elements of the point group. The same can be said about the vectors $\boldsymbol{\rho}_i$ associated with the equivalent atomic positions. For the crystals of the tetrahedral class 23 ($T$) studied in this work, the averaging of the products of vector components associated with the interatomic bonds or atomic positions can be performed as follows
\begin{equation}
\label{eq:amean}
\left< a_\alpha \right> = 0 ,
\end{equation}
\begin{equation}
\label{eq:abmean}
\left< a_\alpha b_\beta \right> = \frac13 (\mathbf{a} \cdot \mathbf{b}) \delta_{\alpha\beta} ,
\end{equation}
\begin{equation}
\label{eq:abcmean}
\begin{split}
\left< a_\alpha b_\beta c_\gamma \right> = & \frac16 \mathbf{a} \cdot [\mathbf{b} \times \mathbf{c}] \varepsilon_{\alpha\beta\gamma} \\ & + \frac16 (|\varepsilon_{\lambda\mu\nu}| a_\lambda b_\mu c_\nu) |\varepsilon_{\alpha\beta\gamma}| ,
\end{split}
\end{equation}
with $\varepsilon_{\alpha\beta\gamma}$ being the antisymmetric Levi-Civita symbol. Using the methods listed above, one can calculate all contributions to the energies $\mathcal{E}^{(2)}$ and $\mathcal{E}^{(3)}$. Some of them are equal to zero or can be reduced to the surface part of the magnetic energy, which is not considered here. In what follows, we list only non-zero contributions, starting from the second order in SOC.
\begin{equation}
\label{eq:E2-1}
\begin{split}
- \frac14 \left.\sum_{i,j}\right.^\prime & J_{ij} \boldsymbol{\mu} \cdot (\tilde{\mathbf{r}}_{ij} \cdot \boldsymbol{\nabla})^2 \boldsymbol{\mu} \\ & = \frac12 \left(\frac16 \left.\sum_{i,j}\right.^\prime J_{ij} \tilde{\mathbf{r}}_{ij}^2 \right) \frac{\partial \boldsymbol{\mu}}{\partial r_\alpha} \cdot \frac{\partial \boldsymbol{\mu}}{\partial r_\alpha} .
\end{split}
\end{equation}
Here we use the averaging $\left< \tilde{r}_{ij,\alpha} \tilde{r}_{ij,\beta} \right> = \frac13 \tilde{\mathbf{r}}_{ij}^2 \delta_{\alpha\beta}$ and the equality $\boldsymbol{\mu} \cdot \Delta \boldsymbol{\mu} = - (\partial \boldsymbol{\mu} / \partial r_\alpha) \cdot (\partial \boldsymbol{\mu} / \partial r_\alpha)$, which is correct up to a total derivative reduceable to a surface integral.
\begin{equation}
\label{eq:E2-2}
\begin{split}
\frac12 \left.\sum_{i,j}\right.^\prime & \mathbf{D}_{ij} \cdot [\boldsymbol{\mu} \times (\tilde{\mathbf{r}}_{ij} \cdot \boldsymbol{\nabla}) \boldsymbol{\mu}] \\ & = \left(- \frac16 \left.\sum_{i,j}\right.^\prime \mathbf{D}_{ij} \cdot \tilde{\mathbf{r}}_{ij} \right)\boldsymbol{\mu} \cdot \mathrm{curl} \boldsymbol{\mu} .
\end{split}
\end{equation}
\begin{equation}
\label{eq:E2-3}
- \mathbf{H} \cdot \left.\sum_i\right.^\prime g_i \mu_\mathrm{B} \cdot \boldsymbol{\mu} = - \mathbf{H} \cdot (M_0 \boldsymbol{\mu}) ,
\end{equation}
where $M_0 \equiv \sum_i^\prime g_i \mu_\mathrm{B}$. The contributions (\ref{eq:E2-1})--(\ref{eq:E2-3}) describe a GL-like energy associated with the continuous magnetization field $\boldsymbol{\mu}(\mathbf{r})$. As a rule, the continuum model of cubic helimagnet is limited to these three terms. In addition to them, there are also two constant contributions associated with antiferromagnetic spin cantings:
\begin{equation}
\label{eq:E2-4}
\frac14 \left.\sum_{i,j}\right.^\prime J_{ij} \left(\mathbf{u}_i^{\prime2} - 2 \mathbf{u}_i^\prime \cdot \mathbf{u}_j^\prime + \mathbf{u}_j^{\prime2}\right) = \frac16 \left.\sum_{i,j}\right.^\prime J_{ij} \left(\boldsymbol{\rho}_i - \boldsymbol{\rho}_j\right)^2 ,
\end{equation}
\begin{equation}
\label{eq:E2-5}
\begin{split}
\frac12 \left.\sum_{i,j}\right.^\prime & \mathbf{D}_{ij} \cdot \left( [\mathbf{u}_i^\prime \times \boldsymbol{\mu}] + [\boldsymbol{\mu} \times \mathbf{u}_j^\prime] \right) \\ & = - \frac13 \left.\sum_{i,j}\right.^\prime \mathbf{D}_{ij} \cdot (\boldsymbol{\rho}_i - \boldsymbol{\rho}_j) \\ & = - \frac23 \left.\sum_i\right.^\prime \boldsymbol{\rho}_i \cdot \sum_j \mathbf{D}_{ij} \\ & = - \frac23 \left.\sum_i\right.^\prime \boldsymbol{\rho}_i \cdot \sum_j J_{ij} \left(\boldsymbol{\rho}_i - \boldsymbol{\rho}_j\right) \\ & = - \frac13 \left.\sum_{i,j}\right.^\prime J_{ij} \left(\boldsymbol{\rho}_i - \boldsymbol{\rho}_j\right)^2 .
\end{split}
\end{equation}
In Eq.~(\ref{eq:E2-5}), we use the symmetry under permutation of the summation indices $i \leftrightarrow j$ and Eq.~(\ref{eq:rho}), which serves as the definition of the rotation vectors $\boldsymbol{\rho}_i$.

Let us also list the nonzero contributions to the energy in the third approximation in SOC.
\begin{equation}
\label{eq:E3-1}
\begin{split}
- \frac12 & \left.\sum_{i,j}\right.^\prime J_{ij} \mathbf{u}_i^{\prime} \cdot (\tilde{\mathbf{r}}_{ij} \cdot \boldsymbol{\nabla}) \mathbf{u}_j^{\prime} \\ & = \frac12 \left.\sum_{i,j}\right.^\prime J_{ij} (\boldsymbol{\rho}_i \cdot (\tilde{\mathbf{r}}_{ij} \cdot \boldsymbol{\nabla}) \boldsymbol{\mu}) (\boldsymbol{\rho}_j \cdot \boldsymbol{\mu}) \\ &
= \left( \frac1{12} \left.\sum_{i,j}\right.^\prime J_{ij} \tilde{\mathbf{r}}_{ij} \cdot [\boldsymbol{\rho}_i \times \boldsymbol{\rho}_j] \right) \boldsymbol{\mu} \cdot \mathrm{curl}\boldsymbol{\mu} ,
\end{split}
\end{equation}
\begin{equation}
\label{eq:E3-2}
\begin{split}
\frac12 & \left.\sum_{i,j}\right.^\prime \mathbf{D}_{ij} \cdot ([\mathbf{u}_i^{\prime} \times (\tilde{\mathbf{r}}_{ij} \cdot \boldsymbol{\nabla}) \boldsymbol{\mu}] + [\boldsymbol{\mu} \times (\tilde{\mathbf{r}}_{ij} \cdot \boldsymbol{\nabla}) \mathbf{u}_j^{\prime}]) \\ & = \frac12 \left.\sum_{i,j}\right.^\prime (\mathbf{D}_{ij} \cdot \boldsymbol{\mu}) (\boldsymbol{\rho}_i + \boldsymbol{\rho}_j) \cdot (\tilde{\mathbf{r}}_{ij} \cdot \boldsymbol{\nabla}) \boldsymbol{\mu} \\ & = \left( \frac1{12} \left.\sum_{i,j}\right.^\prime \mathbf{D}_{ij} \cdot [\tilde{\mathbf{r}}_{ij} \times (\boldsymbol{\rho}_i + \boldsymbol{\rho}_j)] \right) \boldsymbol{\mu} \cdot \mathrm{curl}\boldsymbol{\mu} .
\end{split}
\end{equation}
When deriving Eqs.~(\ref{eq:E3-1}),~(\ref{eq:E3-2}), averaging~(\ref{eq:abcmean}) was used for triples of vectors $\{\tilde{\mathbf{r}}_{ij}, \boldsymbol{\rho}_i, \boldsymbol{\rho}_j\}$ and $\{\mathbf{D}_{ij}, \tilde{\mathbf{r}}_{ ij}, \boldsymbol{\rho}_i + \boldsymbol{\rho}_j\}$, respectively. We neglected the surface terms proportional to $|\varepsilon_{\alpha\beta\gamma}| \mu_\alpha (\partial \mu_\gamma / \partial r_\beta)$. These two terms add up to a correction to the DM parameter of the continuum model. In addition, there is also a constant contribution to the energy
\begin{equation}
\label{eq:E3-3}
\begin{split}
\frac12 \left.\sum_{i,j}\right.^\prime & \mathbf{D}_{ij} \cdot [\mathbf{u}_i^{\prime} \times \mathbf{u}_j^{\prime}] \\ & = \frac12 \left.\sum_{i,j}\right.^\prime (\mathbf{D}_{ij} \cdot \boldsymbol{\mu}) ([\boldsymbol{\rho}_i \times \boldsymbol{\rho}_j] \cdot \boldsymbol{\mu}) \\ & = \frac16 \left.\sum_{i,j}\right.^\prime \mathbf{D}_{ij} \cdot [\boldsymbol{\rho}_i \times \boldsymbol{\rho}_j] .
\end{split}
\end{equation}

It is worth noting that, by the definition~(\ref{eq:mu}) of vector $\boldsymbol{\mu}$, $\sum_i^\prime g_i \mathbf{u}_i = 0$. This means that the energy of interaction with an external magnetic field is zero in the third order in SOC: $-\mathbf{H} \cdot \sum_i^\prime g_i \mu_\mathrm{B} \mathbf{u}_i^{\prime} = 0$.

\end{document}